\documentclass[prl,twocolumn,aps,epsfig,prbbib]{revtex4}
\usepackage{graphicx}

\begin{document}

\title{Averaging of the electron effective mass in 
multicomponent transparent conducting oxides}

\author{Julia E. Medvedeva}\email{juliaem@umr.edu}

\affiliation{Department of Physics, University of Missouri--Rolla, 
Rolla, MO 65409 (USA)}

\begin{abstract}
We find that layered materials composed of various oxides 
of cations with $s^2$ electronic configuration, $XY_2$O$_4$, 
$X$=In or Sc, $Y$=Ga, Zn, Al, Cd and/or Mg, 
exhibit isotropic electron effective mass which can be obtained 
via averaging over those of the corresponding single-cation oxide constituents.
This effect is due to a hybrid nature of the conduction band formed from 
the $s$-states of {\it all} cations and the oxygen $p$-states.
Moreover, the observed insensitivity of the electron effective mass 
to the oxygen coordination and to the distortions in the cation-oxygen chains 
suggests that similar behavior can be expected in 
technologically important amorphous state. 
These findings significantly broaden the range of materials as efficient 
transparent conductor hosts. 
\end{abstract}

%\pacs{71.20.-b}{Electron density of states and band structure of crystalline solids}
%\pacs{72.20.-i}{Conductivity phenomena in semiconductors and insulators}
%\pacs{78.20.Bh}{Optical properties of bulk materials: Theory, models, and numerical simulation}

\maketitle

%\date{\today}

Transparent conducting oxides (TCO) -- the vital part of 
optoelectronic devices --
have been known for a century and employed technologically for decades
\cite{Chopra,Thomas,MRS,TCO-optoelectr,Edwards}. 
Yet, the current TCO market is dominated by only three materials, 
In$_2$O$_3$, SnO$_2$ and ZnO, and the research efforts are 
primarily focused on the oxides of post-transition metals 
with $(n-1)d^{10}ns^2$ electronic configuration.
Despite excellent optical and thermal properties as well as low cost, 
oxides of the main group metals, 
such as Al$_2$O$_3$, SiO$_2$, MgO and CaO,
have never been considered as candidates to achieve useful electrical 
conductivity due to the challenges of efficient carrier generation 
in these wide-bandgap materials \cite{Neumark,VandeWalle,Zunger}.

\begin{figure}
%\centerline{
\includegraphics[width=8.0cm]{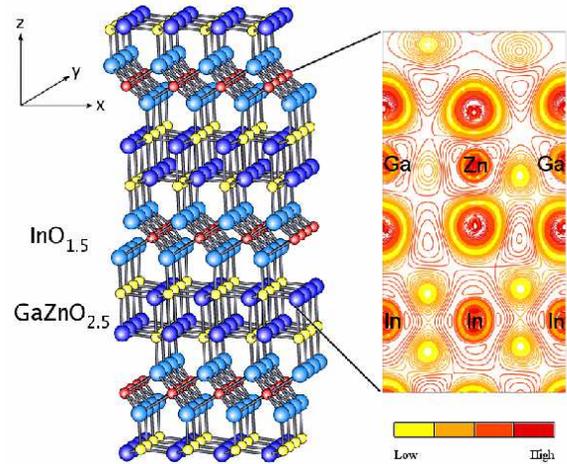}
%}
\caption{The unit cell of InGaZnO$_4$ where three similar blocks 
consisting of one InO$_{1.5}$ and two GaZnO$_{2.5}$ alternate 
along the [0001] direction. Ga and Zn atoms are distributed randomly.
Three-dimensional {\it interatomic} (background) charge density distribution
is evident from the contour plot calculated in the (011) plane.
The plotted charge density corresponds to the carrier concentration 
of $\sim$1$\times$10$^{18}$cm$^{-3}$.
}
\label{cell}
\end{figure}

Multicomponent TCO with layered structure, e.g., InGaZnO$_4$
\cite{Freeman,exp,cluster,PhilMag,Science,Nature},
drew attention due to a possibility to separate carrier donors
(traditionally, oxygen vacancies or aliovalent substitutional dopants)
and the conducting layers where carriers are transfered without charge
scattering on the impurities.
In InGaZnO$_4$, octahedrally coordinated In layers alternate with 
double layers of oxygen tetrahedrons around Ga and Zn, Fig. \ref{cell}.
Because octahedral oxygen coordination of cations was long believed 
to be essential for a good transparent conductor
\cite{Freeman,Shannon,Li-implant,Woodward,Mason-review},
it has been suggested that in InGaZnO$_4$ 
the charge is transfered within the InO$_{1.5}$ layers
while the atoms in GaZnO$_{2.5}$
layers were proposed as candidates for efficient doping 
\cite{Freeman,exp,cluster}. Conversely, it has been argued that InGaO$_3$(ZnO)$_m$
is a Zn 4$s$ conductor \cite{PhilMag}.

To understand the role of local symmetry in the intrinsic transport 
properties of TCO's and to determine the functionality of structurally and 
chemically distinct layers in InGaZnO$_4$,
we employ {\it ab-initio} density functional approach to study 
the electronic properties of various single and multi-cation oxides.
Further, using InGaZnO$_4$ as a test model which examplifies not only the structural 
but also combinatorial peculiarities of complex TCO's, we survey 
other $ns^2$ cations -- beyond the traditional In, Zn and Ga --
for a possibility of being effectively incorporated
into novel multicomponent TCO hosts.

\begin{figure}
%\centerline{
\includegraphics[width=4.0cm]{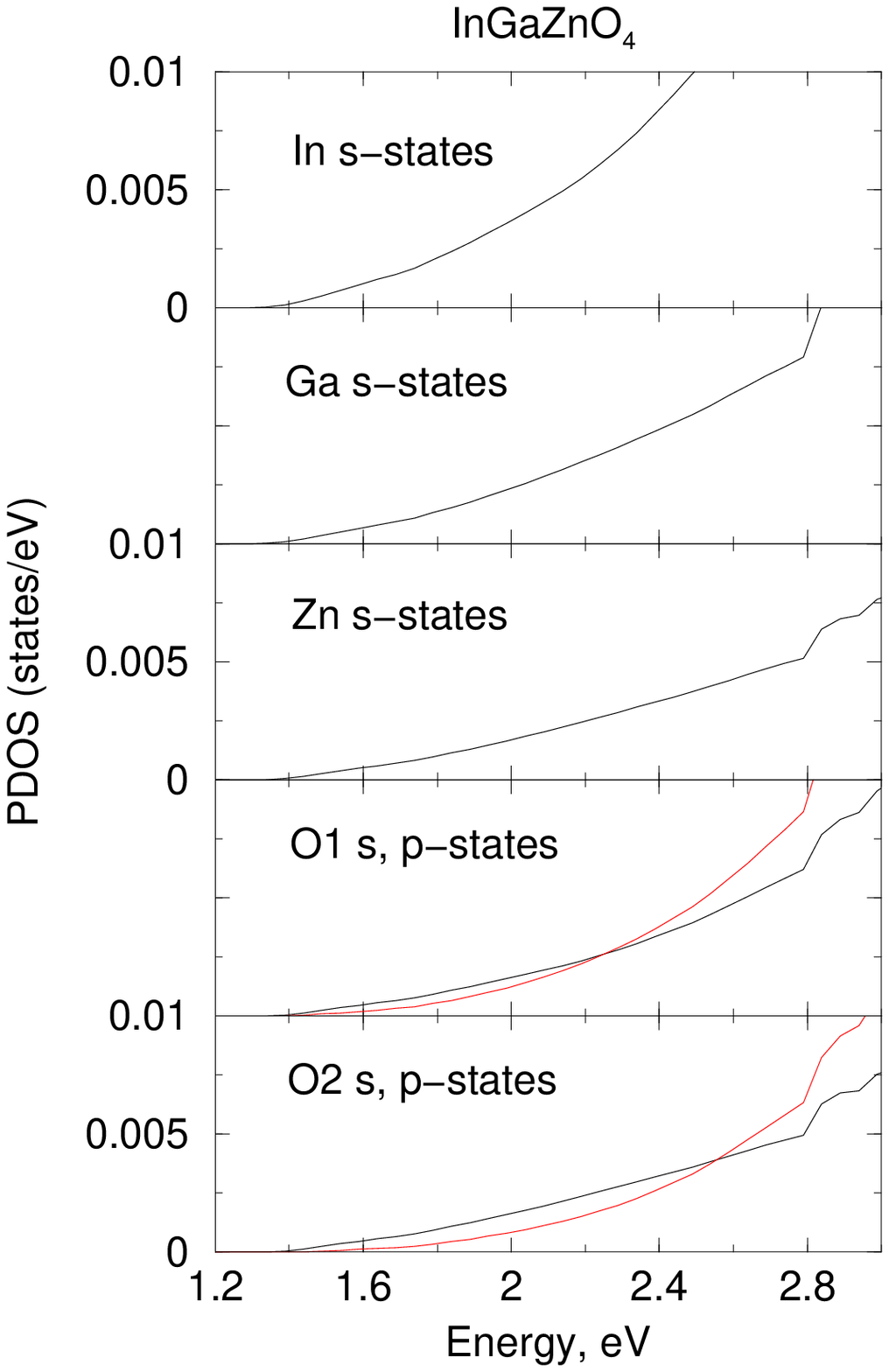}
\includegraphics[width=4.0cm]{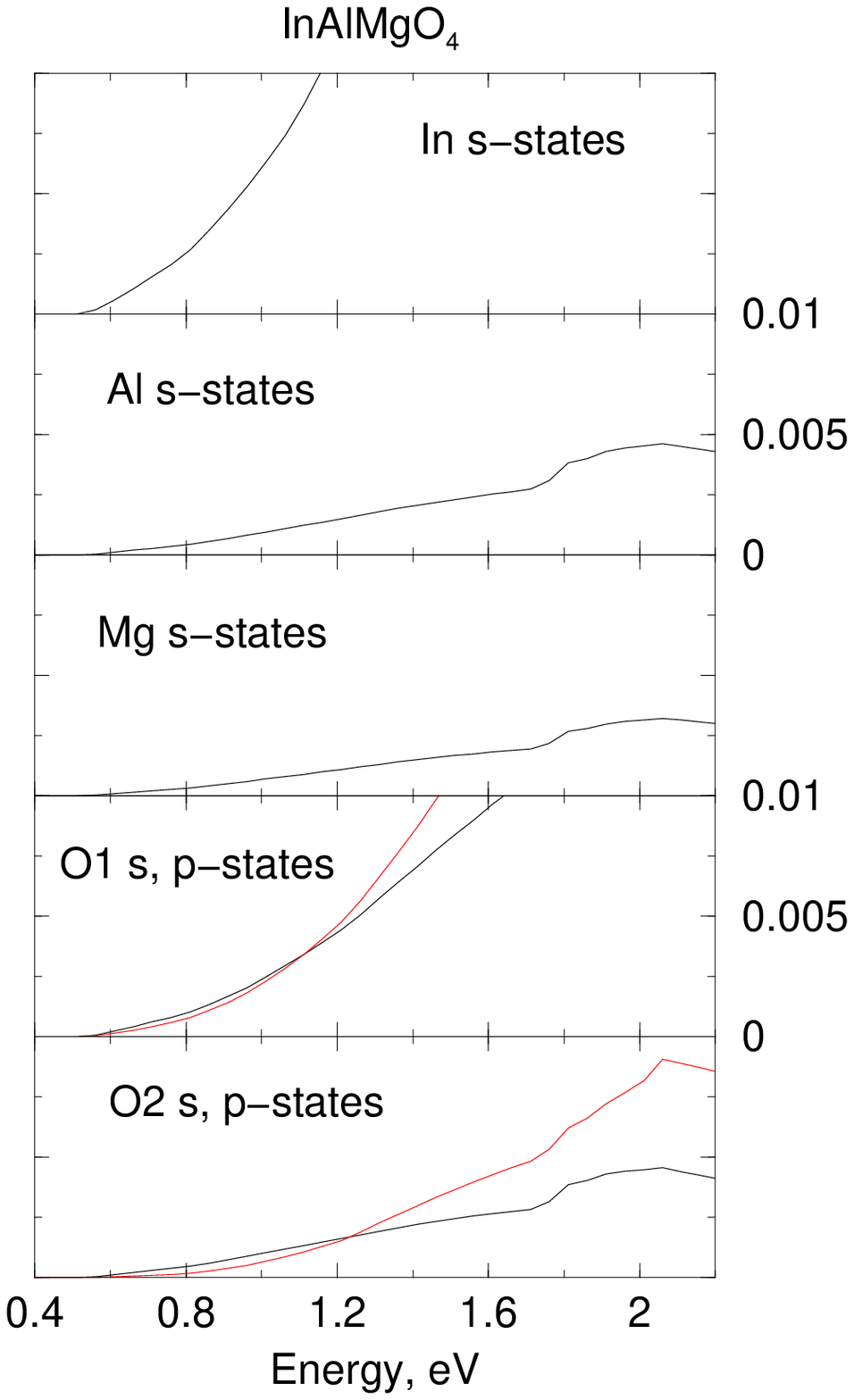}
%}
\caption{
Partial density of states at the bottom of the conduction band for InGaZnO$_4$ and InAlMgO$_4$. 
Atoms from both In-O1 and Ga-Zn-O2 (or Al-Mg-O2) layers give non-negligible contributions.}
\label{pdos}
\end{figure}

{\it Isotropy of the electronic properties in InGaZnO$_4$.}
The electronic band structure calculations for InGaZnO$_4$ show 
that the atoms from both InO$_{1.5}$ and GaZnO$_{2.5}$ layers 
give comparable contributions to the conduction band, fig.~\ref{pdos}, 
leading to a three-dimensional distribution of the charge density, fig.~\ref{cell}.
Moreover, the isotropy of the electronic properties in this layered material
manifests itself in the electron effective masses being nearly 
the same in all crystallographic directions (table~\ref{table-mass}).

\begin{figure*}
%\centerline{
\includegraphics[height=4.7cm]{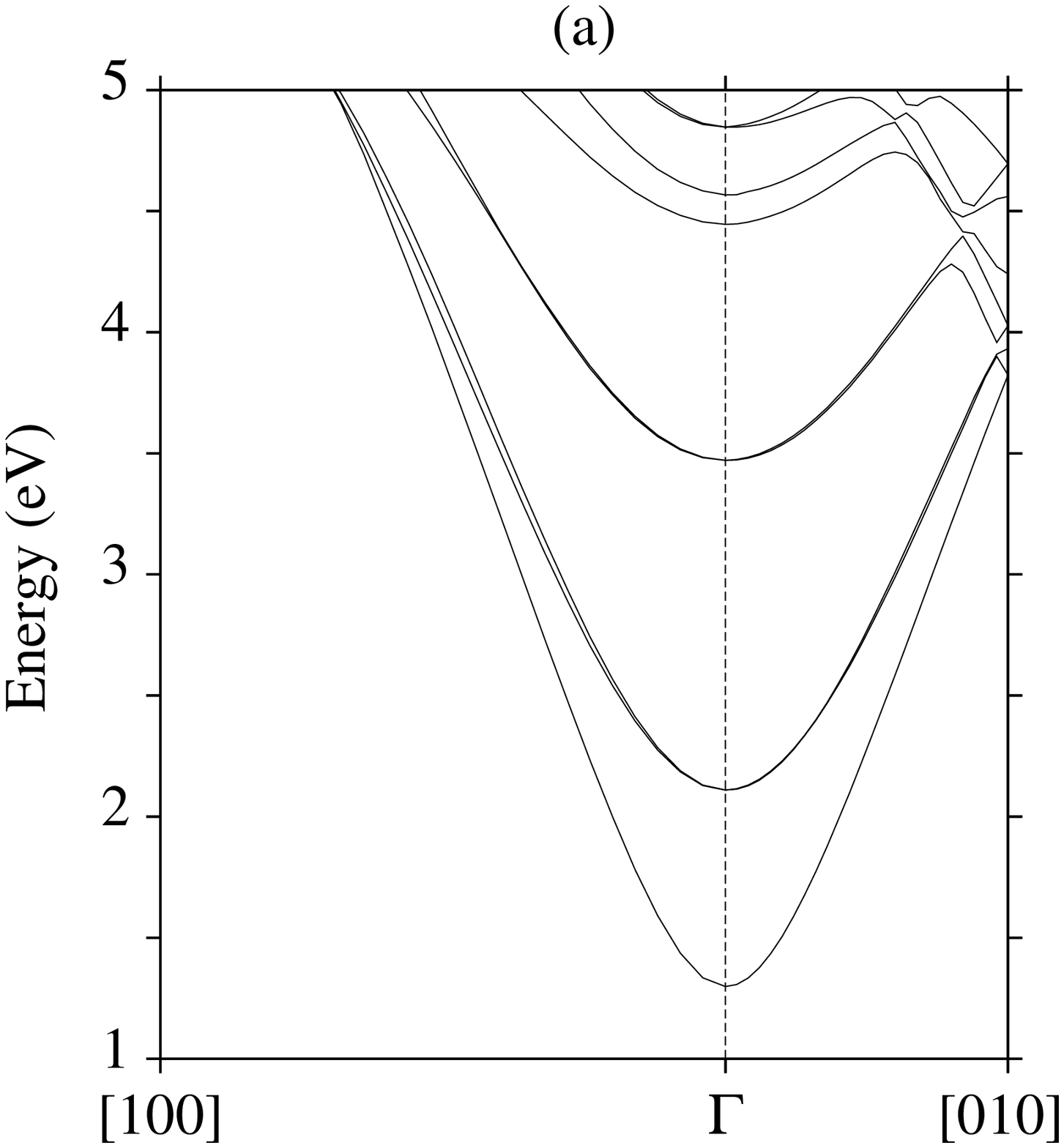}
\includegraphics[height=4.7cm]{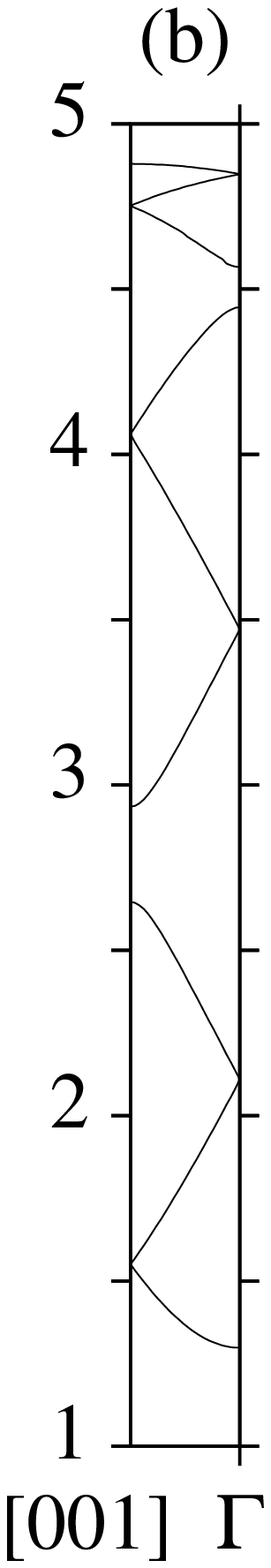}
\includegraphics[height=4.7cm]{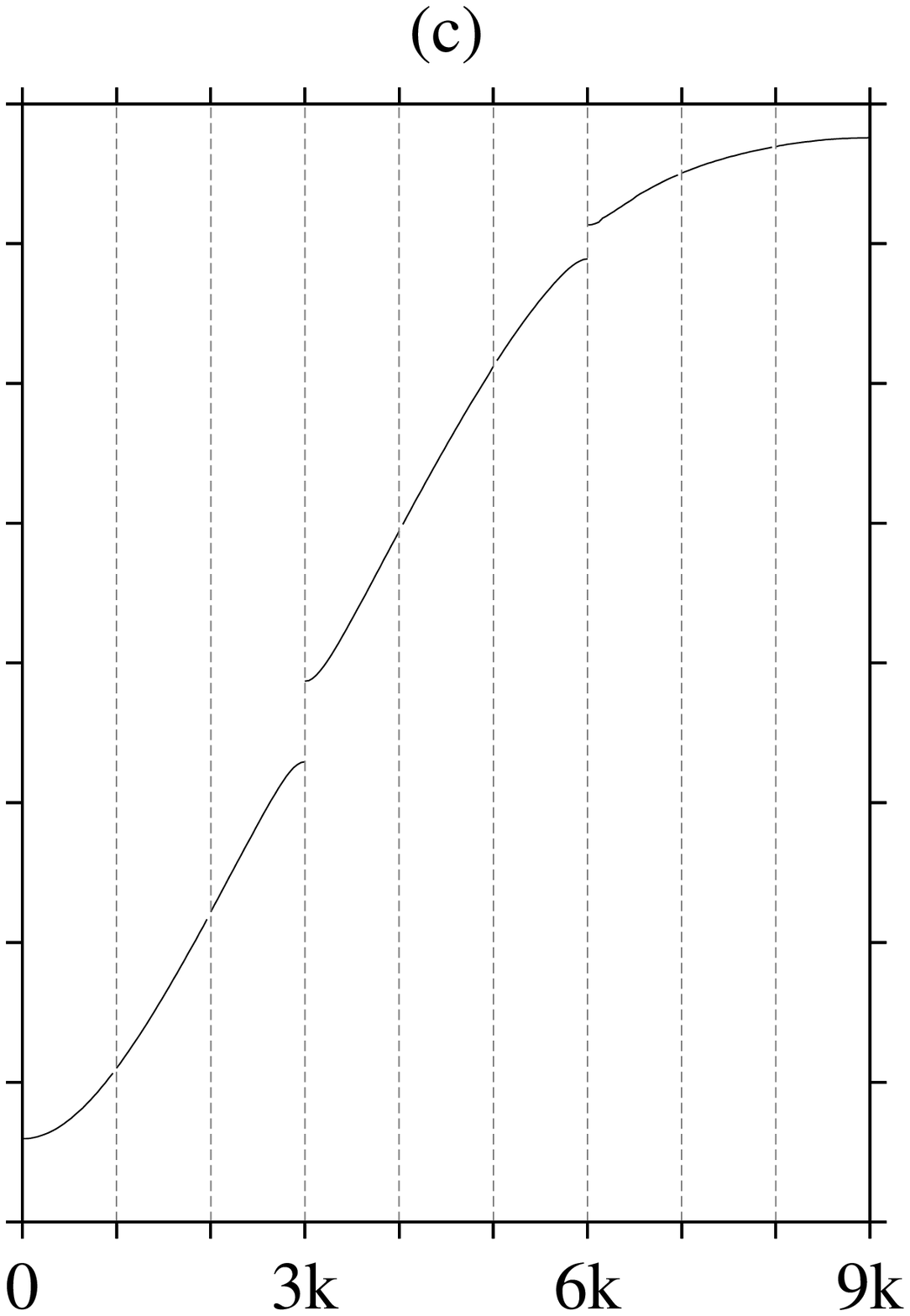}
\includegraphics[height=4.7cm]{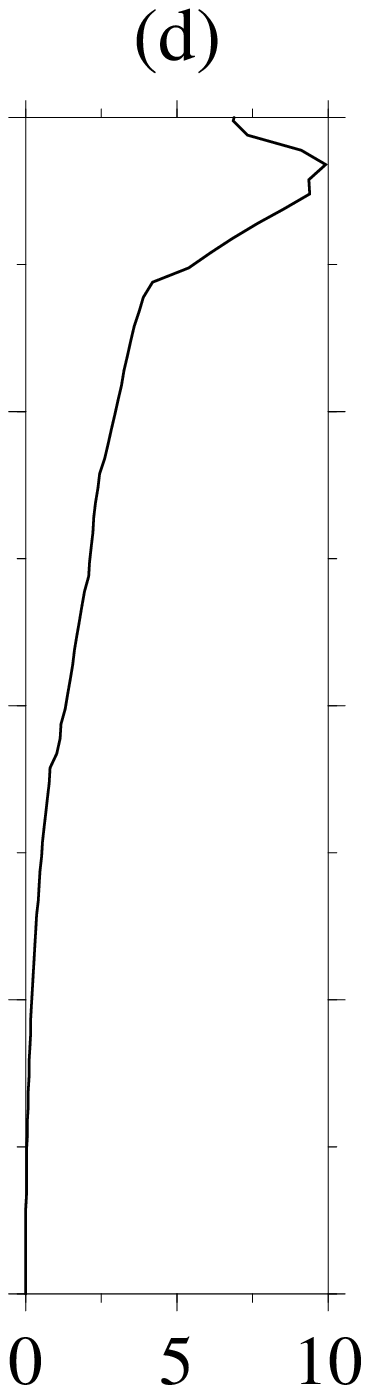}
\includegraphics[height=4.7cm]{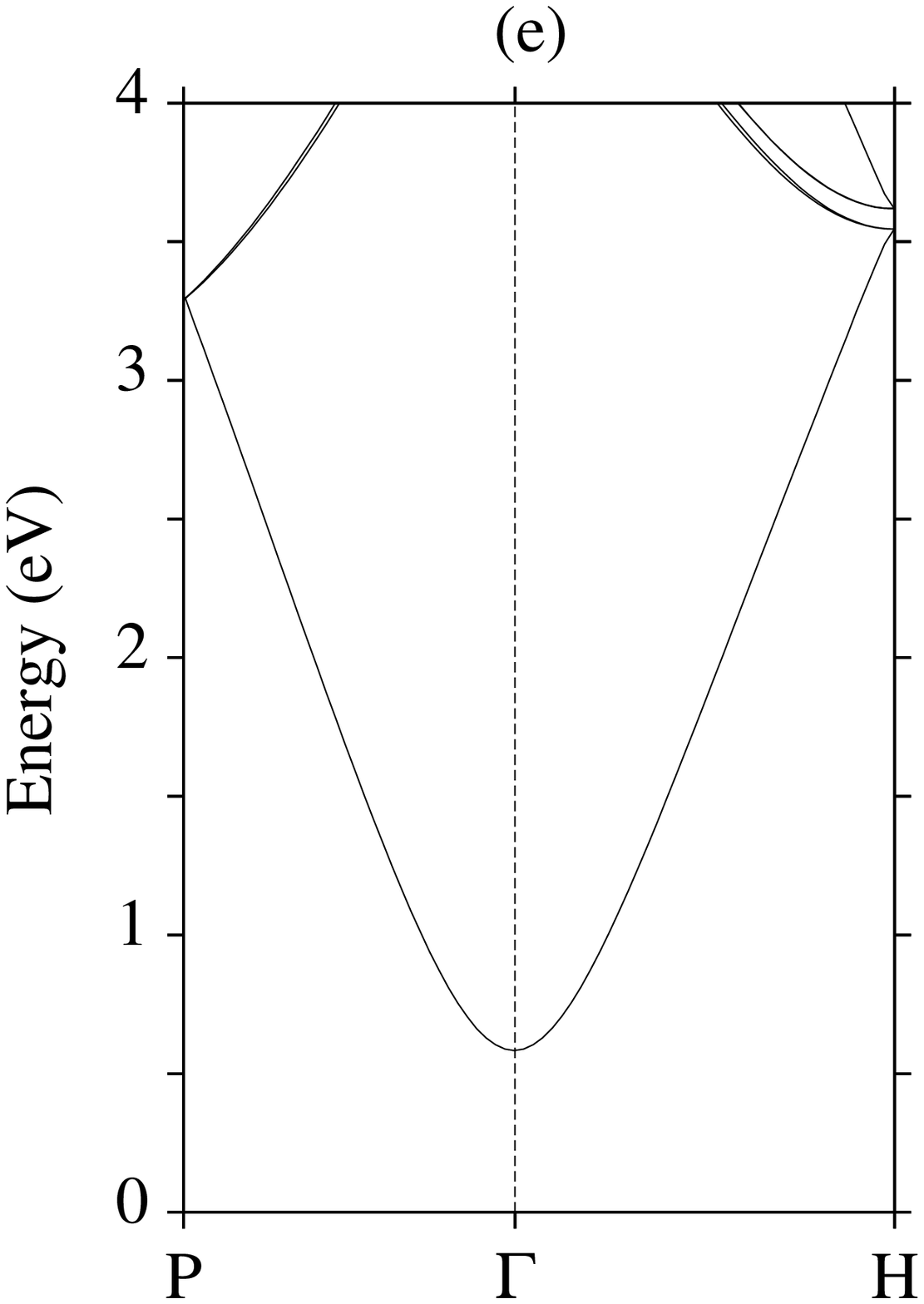}
\includegraphics[height=4.7cm]{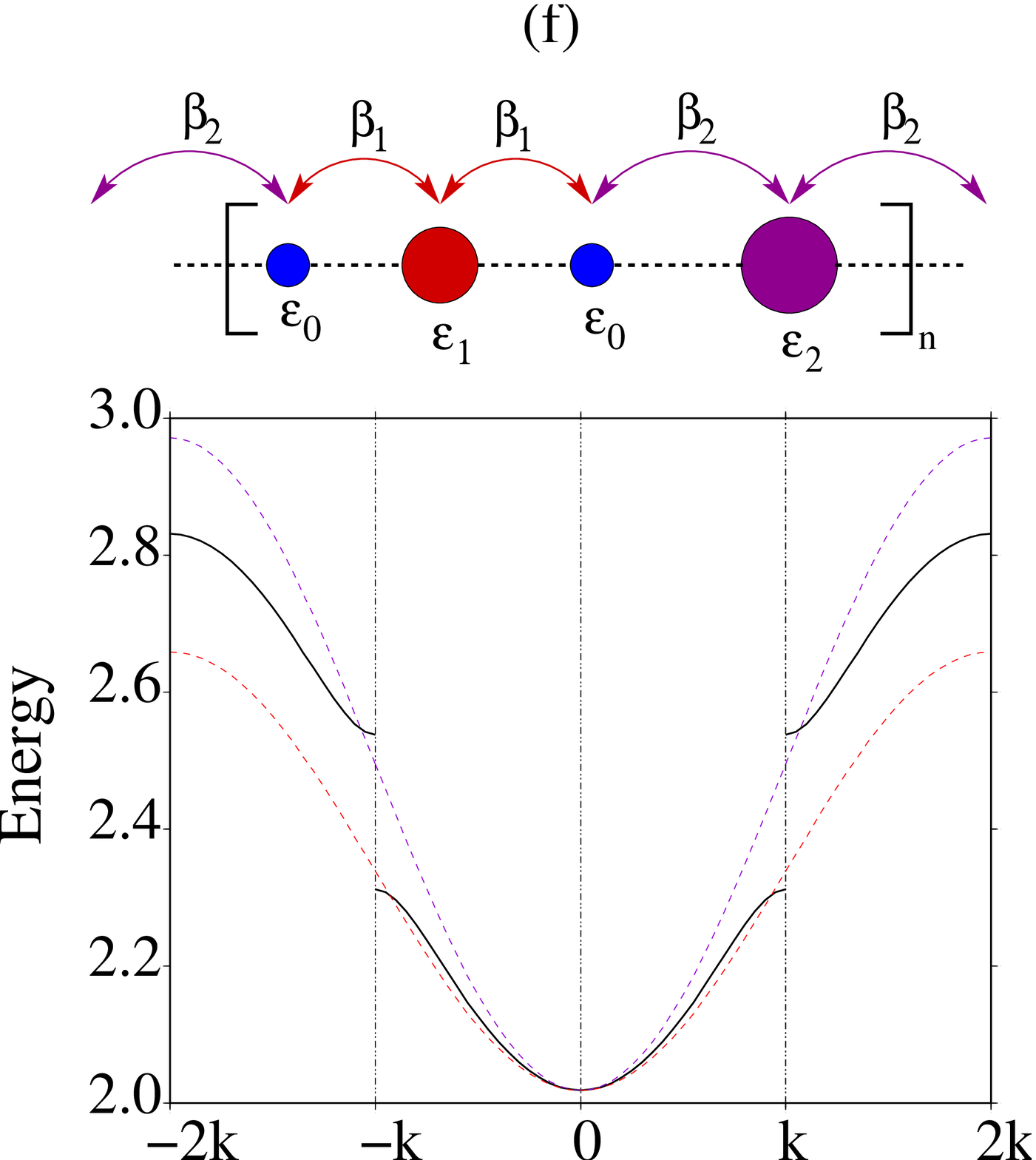}
%}
\caption{Electronic band structure of single and multi-cation oxides. 
(a) Conduction band dispersion calculated in the $ab$-plane and 
(b) along the [0001] direction in InGaZnO$_4$. 
(c) The conduction band unfolded nine times that 
corresponds to the number of [0001] layers in the unit cell,
k=$\frac{\pi}{c}$, $c$ is the lattice parameter. 
The resulting single free-electron-like band is in accord with the stepless 
total density of states (d), in states/eV. 
(e) Conduction band of In$_2$O$_3$ is given for comparison.
(f) Tight-binding conduction band (solid line) calculated for 
one-dimensional atomic chain depicted above the plot. Two types of metal 
atoms (red and purple spheres) alternate with oxygen atoms (blue spheres) 
and only the nearest-neighbor hopping $\beta$ is assumed.
To illustrate the effective mass averaging, cf. eq.~(1), 
the conduction bands for 
the corresponding single-metal oxide chains (dashed lines)
are aligned with $(\varepsilon_1+\varepsilon_2)/2$.}
\label{add-bands}
\end{figure*}

The conduction band in InGaZnO$_4$ consists of a set 
of highly dispersed parabolic bands, fig.~\ref{add-bands}(a).
Since the band gap values in the corresponding single metal oxides 
are different, one may expect that each band is attributed to a certain cation
in this multicomponent compound.
However, we find that each band cannot be assigned to a state of 
a particular atom since all atoms in the cell, including the oxygen 
atoms, give non-negligible contributions to the conduction band 
wavefunction, fig.~\ref{pdos} and table~\ref{table-mass}.  
 
The conduction band dispersion calculated along the [0001] crystallographic 
direction for a {\it single} unit cell, fig.~\ref{add-bands}(b), reveals that 
the multiple bands can be attributed to a ``folding'' of one parent band. 
Triple unfolding of the conduction band corresponds to the three-time reduction 
(expansion) of the conventional unit cell (Brillouin zone) in the $z$ direction.
Since the block of three layers (one InO$_{1.5}$ and two GaZnO$_{2.5}$ layers,
fig.~\ref{cell})
is repeated in the unit cell via translation, the splitting 
between the resulting bands [for the $k$-vector equal to
$\frac{\pi}{c}$, $\frac{2\pi}{c}$, $\frac{4\pi}{c}$, ..., fig.~\ref{add-bands}(c)]
is negligible. 
Although the subsequent unfolding into three individual layers is not 
justified because the three layers are structurally and chemically dissimilar, fig.~\ref{cell},
we find that the band can be unfolded again, fig.~\ref{add-bands}(c). 
The resulting highly dispersed band is in accord 
with the stepless increase of the density of states, fig.~\ref{add-bands}(d). 
Thus, the conduction band can be ``unfolded'' nine times that corresponds to
the total number of layers in the unit cell. Therefore,
the electronic properties of the individual layers are similar.

\begin{table*}
\caption{
Net contributions from the states of the atoms that belong to 
the $X$-O1 or $Y_2$-O2 layers ($X$=In or Sc, $Y$=Ga, Zn, Al, Cd and/or Mg) 
to the conduction band wavefunction at $\Gamma$ point, in per cent; and
the electron effective masses m, in $m_e$,
calculated from the band structure of the layered oxides
and the components of the electron effective-mass tensor, $m_{a,b}$ and $m_z$, 
calculated from eqs.~(1) and (2) using the effective masses of the corresponding 
single-cation oxides.
}
\label{table-mass}
\begin{center}
\begin{tabular}{l|cccc|ccc|cc} \hline
$XY_2$O$_4$ & N$_X$ & N$_{\rm O1}$ & N$_{Y_2}$ & N$_{\rm O2}$ & m$_{[100]}$ & m$_{[010]}$ & m$_{[001]}$ & m$_{ab}$ & m$_z$ \\ \hline
InGaZnO$_4$ & 23 & 25 & 29 & 23 & 0.23 & 0.22 & 0.20 & 0.23 & 0.23 \\
InAlCdO$_4$ & 27 & 27 & 18 & 28 & 0.26 & 0.25 & 0.20 & 0.27 & 0.27 \\
InGaMgO$_4$ & 27 & 31 & 21 & 21 & 0.27 & 0.27 & 0.24 & 0.28 & 0.29 \\
InAlMgO$_4$ & 33 & 40 & 12 & 15 & 0.32 & 0.31 & 0.35 & 0.31 & 0.34 \\
ScGaZnO$_4$ &  8 & 19 & 40 & 33 & 0.33 & 0.33 & 0.34 & 0.33 & 0.53 \\ \hline
\end{tabular}
\end{center}
\end{table*}

{\it Unconventional $s^2$-cations at work.}
A two-dimensional electronic structure could be expected 
in InAlMgO$_4$ and ScGaZnO$_4$ since the band gap values in Sc$_2$O$_3$, 
Al$_2$O$_3$ and MgO are at least twice larger than those in In$_2$O$_3$, 
Ga$_2$O$_3$, CdO and ZnO and, hence, the unoccupied $s$-states of Sc, Al 
and Mg should be located deeper in the conduction band. 
From the analysis of the partial density of states 
for InGaMgO$_4$, InAlCdO$_4$, 
ScGaZnO$_4$ and InAlMgO$_4$, we find that although 
the contributions to the bottom of the conduction band 
from Sc, Al and Mg atoms are notably reduced, these states 
are available for electron transport, fig.~\ref{pdos}. 
Thus, similar to InGaZnO$_4$ where the cations $s$-states are energetically
compatible, in all multicomponent oxides considered, 
the conduction band wavefunction is a combination of 
the $s$-states of {\it all} cations and the $p$-states of the oxygen atoms. 
The contributions from the chemically distinct layers are comparable, 
table~\ref{table-mass}, and, consequently, these complex
oxides exhibit three-dimensional network for the electron transport and
isotropic electron effective mass, table~\ref{table-mass}.

\begin{figure*}
%\centerline{
\includegraphics[width=13.0cm]{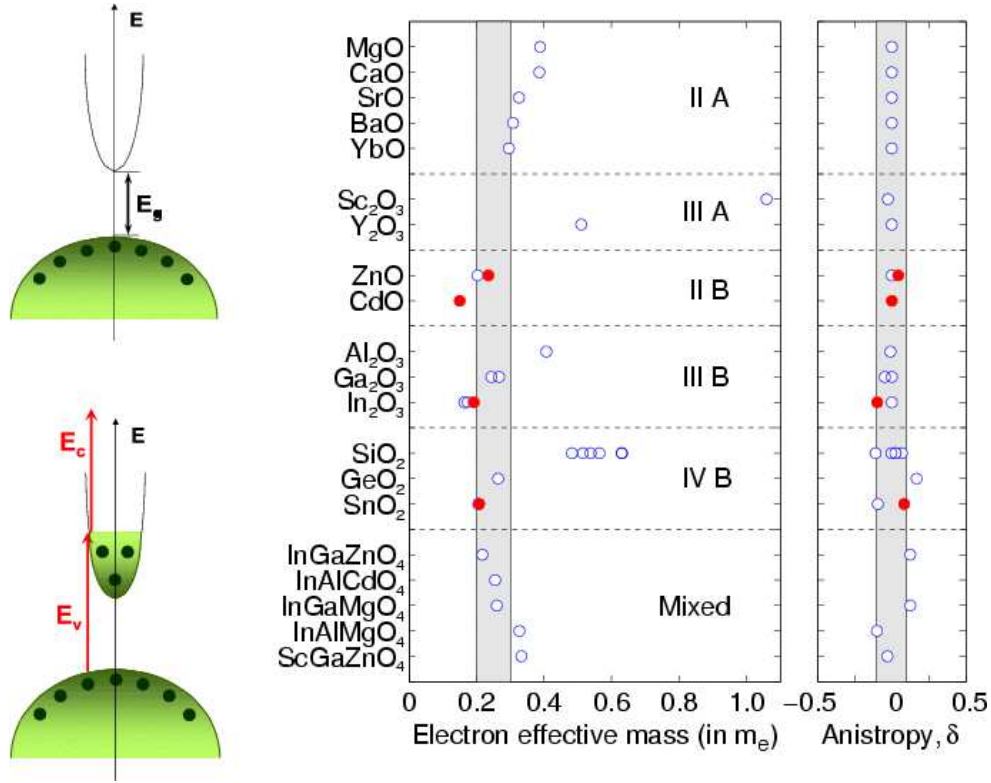}
%}
\caption{
The key electronic features of n-type TCO hosts.
High energy dispersion of the conduction band (left) 
is essential for both high mobility of extra carriers 
due to their small effective mass and low optical absorption. 
For complete transparency in the visible range,
the inter-band transitions E$_v$ and E$_c$ should be $>$3.1 eV, while
the intra-band transitions as well as plasma frequency
should be $<$1.8 eV.
Such a dispersed conduction band along with the above-visible optical 
transitions at the $\Gamma$ point are found in a variety 
of oxides with $s^2$-cation(s). The electron effective mass
calculated for different oxide phases
shows little dependence on the oxygen coordination and 
is isotropic ($\delta$=$(m_a+m_b)/2m_c-1$).
The highlighted areas are to guide the eye; red circles represent currently
known TCO hosts.
}
\label{oxides}
\end{figure*}

{\it Comparison to single-cation TCO's.}
The unfolded conduction band in the layered multicomponent 
materials resembles those of single-cation TCO's,
e.g., In$_2$O$_3$, cf. Figs. \ref{add-bands} (c) and (e).
Such a highly-dispersed single conduction band 
is the key attribute of any conventional \cite{EPL} 
n-type TCO host \cite{Freeman,Woodward,Mryasov,Asahi,JACS,my-PRL}.
Upon proper doping, it provides both high mobility of extra carriers
(electrons) due to their small effective mass, and 
low optical absorption due to high-energy 
inter-band transitions from the valence band, E$_v$, and
from the partially filled conduction band, E$_c$, fig.~\ref{oxides}. 
Even in relatively small bandgap oxides, e.g., CdO where
the optical band gap, E$_g$, is 2.3 eV, the high energy dispersion ensures a
pronounced Fermi energy displacement with doping (so called Burstein-Moss shift)
which helps to keep the intense transitions from the valence band
out of the visible range. However, large carrier concentrations 
required for good electrical conductivity may result in an increase of 
the optical absorption due to low-energy transitions from the Fermi level up 
into the conduction band as well as plasma frequency. 
Application-specific optical properties and desired band offsets 
(work functions) can be attained in a multicomponent transparent conductor
with a proper composition.

In both single and multi-cation oxides, the conduction band is formed from 
the empty $s$-states of the metal atoms and the oxygen antibonding $p$-states, 
e.g., fig.~\ref{pdos}. For multicomponent oxides we find that even at the bottom 
of the conduction band, i.e., at $\Gamma$ point, 
the contributions from the oxygen $p$-states are significant, table~\ref{table-mass}. 
Thus, the key feature of the conduction band in a conventional TCO 
-- its high energy dispersion --
originates in a strong interaction between the cation $s$-states and 
the anion antibonding $p$-states \cite{Woodward}. The direct 
$s$-$s$ overlap is insignificant, fig.~\ref{cell}, and therefore,
the $s$-$s$ interactions which were commonly assumed to play to key role 
in the electronic properties of TCO \cite{Li-implant,PhilMag,Nature},
do not govern the transport properties in these oxides.  
Indeed, the small electron effective mass in $s^2$-cation oxides
is determined by the strong $s$-$p$ interactions \cite{kp}.

From orbital symmetry considerations, one can see that 
the oxygen coordination {\it of cations} does not affect the 
$s$-$p$ overlap. Instead,
the largest overlap should be attained in materials where the oxygen atom 
is coordinated octahedrally {\it by cations} with the extended $s$-orbitals \cite{Woodward}.
Our systematic comparison of the calculated electron effective mass 
in the oxides of metals with $s^2$ electronic configuration, fig.~\ref{oxides},
shows that the mass slightly decreases as the ionic radii of the cations from 
the same group or the symmetry of the same-cation oxide increases.
However, variations in the oxygen coordination as well as strong 
distortions in the metal-oxygen chains in different oxide phases
lead to insignificant changes in the effective mass.
For example, for cubic (octahedral coordination) and hexagonal 
(tetrahedral) ZnO or
for corundum (distorted tetrahedral) and monoclinic $\beta$-phase 
(both distorted tetrahedral and trigonal) Ga$_2$O$_3$
the corresponding electron effective masses vary by 15\% or 9\%, 
respectively. The largest deviation in the effective mass values  
for various SiO$_2$ phases is 26\%.
Furthermore, the effective mass remains isotropic for all phases of 
the $s^2$-cation oxides -- including silica ITQ-4 zeolite with large pore channels 
\cite{exception}.
These observations explain the success of amorphous transparent conducting oxides 
-- in marked contrast to the amorphous Si where the directional interactions between 
the conduction $p$-orbitals lead to strong anisotropy of the transport properties 
which are sensitive to the orbital overlap and hence to the distortions 
in the atomic chains \cite{Nature}.
Finally, we note that the fact that the calculated as well as the observed \cite{expSnO2} 
isotropic effective mass in rutile SnO$_2$ 
where the direct overlap between Sn $s$-orbitals is possible (only)
along the [001] direction, corroborate our conclusion that the $s$-$s$ interactions  
do not govern the transport properties as discussed above.

{\it Effective mass averaging.}
Because the local symmetry (nearest neighbors), 
the cation-oxygen bond lengths and, hence, the $s$-$p$ overlap 
are similar in the single and multi-cation oxides, 
the intrinsic transport properties in the layered materials should be related
to those in the single-cation oxides.
Moreover, due to the hybrid nature of the conduction states in the multicomponent oxides, 
the states of all cations should give the same order of magnitude contributions to 
the effective mass. Thus, we expect the later to be an ``effective'' average over 
the effective masses of the corresponding single-cation oxides. 
Indeed, we find that in case of one-dimensional atomic chain where two types 
of metal atoms alternate with oxygen atoms, the effective mass averaging 
can be derived analytically, fig.~\ref{add-bands}(f).

We formulate a simple approach which allows one to estimate the effective mass of 
the multicomponent oxides as follows.
With proper doping, the Fermi energy is shifted up into 
the conduction band (Burstein-Moss shift).
When the extra electrons propagate along the $z$ direction, 
i.e., across the layers, 
the resulting resistivity is a sum of the resistivities of each layer.
Therefore, the $z$ component of the average effective-mass tensor 
can be found as:
\begin{equation}
m _z = (m_1+m_2+m_3)/3,
%\DeclareMathAlphabet{\mathpzc}{OT1}{pzc}{m}{it}
\end{equation}
where $m_{1,2,3}$ are the effective masses of the corresponding single metal
oxides -- In$_2$O$_3$, Ga$_2$O$_3$ and ZnO in the case of InGaZnO$_4$.

For the in-plane charge transport in the layered materials, the effective-mass tensor 
components can be found in a parallel fashion. 
Note, that one needs to average the effective mass for the mixed 
GaZnO$_{2.5}$ layers:
\begin{equation}
\frac{1}{ {m}  _{a,b}} = \frac{1}{3} \left( \frac{1}{m_1}+\frac{2}{\frac{1}{2}(m_2+m_3)} \right). 
\end{equation}
The resulting $m_{a,b}$ and $m_z$ are presented in table~\ref{table-mass}.
We find that the increase of the electron effective masses
in the order 
InGaZnO$_4$ $<$ InAlCdO$_4$ $<$ InGaMgO$_4$ $<$ InAlMgO$_4$ $<$ ScGaZnO$_4$
is well reproduced by the above averaging.
Moreover, the $m_{a,b}$ and $m_z$ values nearly coincide with 
the corresponding effective masses of the multi-cation oxides 
with the exception of the Sc case. In fact, for 
the Sc-containing compounds the effective mass averaging is not 
legitimate due to the presence of the empty $d$-states of Sc
near the conduction band edge.
In Sc$_2$O$_3$, the Sc $d$-states are located at 0.5 eV above
the conduction band edge, and therefore give significant 
contributions to the effective mass which is about the mass of an electron. 
In the multi-cation oxide, the Sc $d$-states are found to be 
at $\sim$2 eV above the conduction band edge, so that the resulting 
small effective mass, table~\ref{table-mass}, is determined primarily 
by the $s-p$ interactions -- similar to the rest of the $s^2$-cation oxides.

The effective mass averaging procedure, eqs.~(1) and (2),
can be generalized for materials 
consisting of any number of layers, e.g., for InGaO$_3$(ZnO)$_m$, $m$=integer.
Furthermore, since the intrinsic transport properties are determined entirely
by the local symmetry, i.e., the nearest neighbors, 
the effective mass averaging should apply to TCO's
in amorphous state. In this case, one needs to average 
the components of the effective-mass tensor, $m_{amorph}$=$(m_a$+$m_b$+$m_z)/3$.

{\it Importance of carrier generation.}
Upon doping of a TCO host material, the resulting conductivity depends not only on 
the effective mass but also on the carrier generation mechanism, 
carrier concentration and carrier relaxation time. 
Doping of a structurally anisotropic material may lead
to non-uniform distribution of the carrier donors and, therefore, 
the isotropic behavior of the host may not be maintained as, for example, 
in oxygen-deficient $\beta$-Ga$_2$O$_3$ \cite{betaGa2O3}.
In multicomponent InGaO$_3$(ZnO)$_m$, different valence states (In$^{3+}$ and Ga$^{3+}$ 
vs Zn$^{2+}$) and oxygen coordination (octahedral for In vs tetrahedral for Ga and Zn) 
are likely to result in preferential arrangement of 
aliovalent substitutional dopants or oxygen vacancies.
Consequently, an anisotropic mobility should be expected in the layered materials
due to the spatial separation of the carrier donors and the layers where 
the extra carriers are transfered efficiently, i.e., without charge scattering
on the impurities. 
While targeted doping can help make either or both structurally distinct 
layers conducting, the amorphous complex oxides readily offer a way to maintain 
isotropic transport properties. Indeed, experimental observations that the mobility 
and conductivity 
are independent of the large variations in the composition in amorphous 
InGaO$_3$(ZnO)$_n$ with $n$$\leq$4 \cite{PhilMag} and
that the effective masses of amorphous and crystalline InGaZnO$_4$
are nearly the same \cite{a-mass} support our conclusions.

For efficient doping of wide-bandgap oxides
such as MgO, CaO, SiO$_2$ and Al$_2$O$_3$, 
novel carrier generation mechanisms should be sought.
A non-traditional approach has already yielded promising results
in calcium aluminates \cite{mayenite,may-science,PRLmay} -- a conceptually new class of
transparent conductors \cite{EPL}.
Multicomponent oxides -- such as those considered in this work, 
ordered ternary oxides \cite{Minami} as well as their
solid solutions and amorphous counterparts -- 
represent an alternative way to utilize 
the abundant main-group elements such as Ca, Mg, Si and Al
towards novel TCO hosts with a predictable effective mass and optical 
and transport properties controllable via the composition.

{\bf Acknowledgement.} 
The work is supported by University of Missouri Research Board.

%\newpage

\begin{footnotesize} 

APPENDIX

{\it Theoretical methods}
First-principles full-potential linearized augmented plane wave method 
\cite{FLAPW,FLAPW1} 
with the local density approximation is employed for electronic band structure
investigations of the $XY_2$O$_4$ compounds, $X$ = In or Sc and $Y$ = Ga, Zn, Al, 
Cd and/or Mg, and single-cation oxides.
Cut-offs of the plane-wave basis 
(16.0 Ry) and potential representation (81.0 Ry), and expansion in terms 
of spherical harmonics with $\ell \le$ 8 inside the muffin-tin spheres 
were used. Summations over the Brillouin zone were carried out 
using 14 special {\bf k} points in the irreducible wedge.

{\it Structure optimization.}
$XY_2$O$_4$ compounds have rhombohedral $R\bar{3}m$ crystal 
structure of YbFe$_2$O$_4$ type \cite{str,str-all}.
Indium (or scandium)  atoms substitute Yb in 3(a) position, while
both $Y^{3+}$ and $Y^{2+}$ atoms replace Fe in 6(c) position and
are distributed randomly \cite{Li}.
Our total energy calculations for several structures in the ($a$,2$a$,$c$) 
supercell with various arrangements of the $Y^{2+}$ 
and $Y^{3+}$ atoms suggest that their arrangement is not ordered but random -- 
in agreement with the experiment. 
We note here that the electronic band structure features are similar 
for different spatial distributions of the $Y$ atoms for the reasons
discussed in the paper. 

Since the valence state and ionic radii of $Y^{2+}$ and $Y^{3+}$  
are different, the site positions of these atoms as well as their oxygen 
surrounding should be different. 
Because the exact internal positions of atoms are unknown, 
we used those of the YbFe$_2$O$_4$ \cite{str} as the starting values 
and then optimized each structure via the total energy and atomic 
forces minimization. During the optimization, the lattice parameters 
were fixed at the experimental values \cite{str-all,family}.
We find that the optimized cation-anion distances 
correlate with ionic radii of the cations.

{\it Single-cation oxides.}
For the single-cation oxides, the following phases have been calculated:
$Fm\bar{3}m$ for MgO, CaO, SrO, BaO and YbO; $Ia\bar{3}$ for Sc$_2$O$_3$ 
and Y$_2$O$_3$; $Fm\bar{3}m$ and $P6_3mc$ for ZnO; $Fm\bar{3}m$ for CdO;
$R\bar{3}c$ for Al$_2$O$_3$; $R\bar{3}c$ and $C2/m$ for Ga$_2$O$_3$;
$Ia\bar{3}$, $R\bar{3}c$ and $I2_13$ for In$_2$O$_3$; 
$P3_221$, $P2_1/c$, $P6_3/mmc$, $P4_12_12$
and $I2/m$ for SiO$_2$;
$P4_12_12$ and $P4_2/mnm$ for GeO$_2$; $P4_2/mnm$ and $Pbcn$ for SnO$_2$.
For each structure, the internal positions of all atoms have been optimized 
via the total energy and atomic forces minimization, while the lattice parameters 
were fixed at the experimental values.

{\it One-dimensional model of complex oxides.}
The effective mass averaging, cf. eq.~(1), can be shown analytically 
using a one-dimensional model in the tight-binding approximation. 
To capture the key features of complex oxides, we consider a chain consisting 
of two types of metal atoms which alternate with oxygen atoms, 
fig.~\ref{add-bands}(f), and assume only the nearest-neighbor interactions 
given by the hopping integrals $\beta_1$ and $\beta_2$. The Hamiltonian
of this model system is:
\begin{equation}
H=\sum_{n,l} |n,l \rangle \varepsilon_l \langle n,l| + 
\sum_{n,n',l,l'} |n',l' \rangle \beta_l \langle n,l|.
\end{equation}
Here, $l$ is the atom index in the unit cell, $n$ enumerates the cells and 
$n',l'$ in the second sum run over the nearest neighbors.
For the bottom of the conduction band, the dispersion relation can be simplified to
\begin{equation}
\varepsilon(k)=\frac{\varepsilon_1+\varepsilon_2}{2}+
\frac{1}{\frac{1}{2}(\frac{\Delta}{\beta_1^2}+\frac{\Delta}{\beta_2^2})}\left[ ka \right]^2
\end{equation}
if $|\varepsilon_1-\varepsilon_2|<2\left|\frac{\beta_1^2-\beta_2^2}{\Delta}\right|$.
Here $\varepsilon_0$, $\varepsilon_1$ and $\varepsilon_2$ are the atomic level energies
of the oxygen and two types of metal atoms, respectively, and it is assumed that  
$\varepsilon_0 < \varepsilon_{1,2}$ and $\varepsilon_1 \sim \varepsilon_2$;
$\Delta=\frac{1}{2}(\varepsilon_1+\varepsilon_2) - \varepsilon_0$ and $a$ is a half 
of the lattice parameter. 
Similar considerations for the chain consisting of only one type of metal atoms
alternating with oxygen atoms show that the quantity $\frac{\Delta}{\beta^2}$ 
represents the effective mass of the system. Therefore, eq.~(4) represents 
the effective mass averaging over those of the corresponding single-metal ``oxide'' chains,
cf., fig.~\ref{add-bands}(f) 
-- in agreement with the results of our first-principles calculations.
The following parameters were used to plot fig.~\ref{add-bands}(f): 
$\varepsilon_0$ = 1.00, $\varepsilon_1$ = 2.00, $\varepsilon_2$ = 2.05, $\beta_1$ = 0.4 and
$\beta_2$ = 0.5.

\end{footnotesize}


\begin{thebibliography}{99}
\bibitem{Chopra}
K.L. Chopra, S. Major, D.K. Pandya,  
%Transparent conductors -- A status review. 
{\it Thin Solid Films} {\bf 1983}, {\it 102}, 1.
%-46.
\bibitem{Thomas}
G. Thomas, 
%Invisible circuits. 
{\it Nature} {\bf 1997}, {\it 389}, 907.
%-908.
\bibitem{MRS}
Special issue on Transparent Conducting Oxides, 
%D.S. Ginley and C. Bright, 
{\it MRS Bull.} {\bf 2000}, {\it 25}.
%\bibitem{Wager}
%J.F. Wager, 
%Transparent electronics. 
%{\it Science} {\bf 300}, 1245
%-1246 
%(2003).
\bibitem{TCO-optoelectr}
H. Ohta, H. Hosono,   
%Transparent oxide optoelectronics. 
{\it Mater. Today} {\bf 2004}, {\it 7}, 42.
%-51 (2004).
\bibitem{Edwards}
P.P. Edwards, A. Porch, M.O. Jones, D.V. Morgan, R.M. Perks,  
%Basic materials physics of transparent conducting oxides.
{\it Dalton Trans.} {\bf 2004}, {\it 19}, 2995.
%-3002 (2004).

\bibitem{Neumark}
G.F. Neumark, 
%Defects in wide band gap II-VI crystals.
{\it Mater. Sci. Eng. R} {\bf 1997}, {\it 21}, 1.
%-46 (1997). 
\bibitem{VandeWalle}
C.G. Van de Walle,  
%Strategies for controlling the conductivity of wide-band-gap semiconductors.
{\it Phys. Stat. Solidi B} {\bf 2002}, {\it 229}, 221.
%-228 (2002).
\bibitem{Zunger}
A. Zunger,  
%Practical doping principles. 
{\it Appl. Phys. Lett.} {\bf 2003}, {\it 83}, 57.
%-59 (2003).


\bibitem{Freeman}
A.J. Freeman, K.R. Poeppelmeier, T.O. Mason, R.P.H. Chang, T.J. Marks, 
%Chemical and Thin-Film Strategies for New Transparent Conducting Oxides.
{\it MRS Bull.} {\bf 2000}, {\it 25}, 45.
%-51 (2000).
\bibitem{exp}
M. Orita, M. Takeuchi, H. Sakai, H. Tanji, 
%New Transparent Conductive Oxides with  YbFe$_2$O$_4$ Structure.
{\it Jpn. J. Appl. Phys.} {\bf 1995}, {\it 34}, L1550.
%-L1552 (1995).
\bibitem{cluster}
M. Orita, H. Tanji, M. Mizuno, H. Adachi, I. Tanaka, 
%Mechanism of electrical conductivity of transparent InGaZnO$_4$.
{\it Phys. Rev. B} {\bf 2000}, {\it 61}, 1811.
%-1815 (2000).
\bibitem{PhilMag}
M. Orita, H. Ohta, M. Hirano, S. Narushima, H. Hosono, 
%Amorphous transparent conductive oxide InGaO$_3$(ZnO)$_{(m)}$ ($m$$\leq$4): 
%a Zn 4s conductor.
{\it Phil. Mag. B} {\bf 2001}, {\it 81}, 501.
%-515 (2001).
\bibitem{Science}
K. Nomura, H. Ohta, K. Ueda, T. Kamiya, M. Hirano, H. Hosono, 
%Thin-film transistor fabricated in single-crystalline transparent oxide 
%semiconductor.
{\it Science} {\bf 2003}, {\it 300}, 1269.
%-1272 (2003).
\bibitem{Nature}
K. Nomura, H. Ohta, A. Takagi, T. Kamiya, M. Hirano, H. Hosono,
%Room temperature fabrication of transparent flexible thin-film transistors
%using amorphous oxide semiconductors.
{\it Nature} {\bf 2004}, {\it 432}, 488.
%-492 (2004).

\bibitem{Shannon}
R.D. Shannon, J.L. Gillson, R.J. Bouchard,  
%Single crystal synthesis and electrical properties of CdSnO$_3$, 
%Cd$_2$SnO$_4$, In$_2$TeO$_6$, CdIn$_2$O$_4$.
{\it J. Phys. Chem. Solids} {\bf 1977}, {\it 38}, 877.
%-881 (1977).
\bibitem{Li-implant}
H. Kawazoe, N. Ueda, H. Un'no, T. Omata, H. Hosono, H. Tanoue,
%Generation of electron carriers in insulating thin film of MgIn$_2$O$_4$ 
%spinel by Li$^+$  implantation.
{\it J. Appl. Phys.} {\bf 1994}, {\it 76}, 7935.
%-7941 (1994).

\bibitem{Woodward}
H. Mizoguchi, P.M. Woodward, 
%Electronic structure studies of main group oxides possessing edge-sharing octahedra: 
%Implications for the design of transparent conducting oxides. 
{\it Chem. Mater.} {\bf 2004}, {\it 16}, 5233.
%-5248 (2004). 
\bibitem{Mason-review}
B.J. Ingram, G.B. Gonzalez, D.R. Kammler, M.I. Bertoni, T.O. Mason,
%Chemical and structural factors governing transparent conductivity in oxides
{\it J. Electroceram.} {\bf 2004}, {\it 13}, 167.
%-175 (2004).


\bibitem{EPL}
J.E. Medvedeva, A.J. Freeman, 
%Combining high conductivity with complete optical transparency: A band-structure approach.
{\it Europhys. Lett.} {\bf 2005}, {\it 69}, 583.
%-587 (2005). 
\bibitem{Mryasov}
O.N. Mryasov, A.J. Freeman, 
%Electronic band structure of indium tin oxide and criteria for transparent conducting behavior. 
{\it Phys. Rev. B} {\bf 2001}, {\it 64}, 233111.
\bibitem{Asahi}
R. Asahi, A. Wang, J.R. Babcock, N.L. Edleman, A.W. Metz, M.A. Lane,  
V.P. Dravid, C.R. Kannewurf, A.J. Freeman, T.J. Marks,  
%First-principles calculations for understanding high conductivity 
%and optical transparency in In$_x$Cd$_{1-x}$O films.
{\it Thin Solid Films} {\bf 2002}, {\it 411}, 101.
%-105 (2002).
\bibitem{JACS}
Y. Yang, S. Jin, J.E. Medvedeva, J.R. Ireland, A.W. Metz, J. Ni,  
M.C. Hersam, A.J. Freeman, T.J. Marks,  
%CdO as the Archetypical Transparent Conducting Oxide. Systematics of 
%Dopant Ionic Radius and Electronic Structure Effects on Charge Transport and Band Structure
{\it J. Amer. Chem. Soc.} {\bf 2005}, {\it 127}, 8796.
%-8804 (2005).
\bibitem{my-PRL}
J.E. Medvedeva, 
%Magnetically mediated transparent conductors: In$_2$O$_3$ doped with Mo.
{\it Phys. Rev. Lett.} {\bf 2006}, {\it 97}, 086401.

\bibitem{exception}
One exception is SiO$_2$ in high-temperature rutile phase 
with two unique Si-O bonds: the calculated effective mass 
in the $ab$-plane (long Si-O bond) is 2.6 times larger than the one 
in the $z$ direction (short Si-O bond).

\bibitem{expSnO2}
K.J. Button, D.G. Fonstad, W. Dreybradt, {\it Phys. Rev. B} {\bf 1971}, {\it 4}, 4539.

\bibitem{kp}
C. Kittel, {\it Introduction to Solid State Physics}, p. 253 (John Wiley and Sons, Inc., 2005).

\bibitem{betaGa2O3}
M. Yamaga, E.G. Villora, K. Shimamura, N. Ichinose, M. Honda, {\it Phys. Rev. B}
{\bf 2003}, {\it 68}, 155207. 


\bibitem{a-mass}
A. Takagi, K. Nomura, H. Ohta, H. Yanagi, T. Kamiya, M. Hirano, H. Hosono,
%Carrier transport and electronic structure in amorphous oxide semiconductor, a-InGaZnO$_4$
{\it Thin Solid Films} {\bf 2005}, {\it 486}, 38.
%-41 (2005).


\bibitem{mayenite}
K. Hayashi, S. Matsuishi, T. Kamiya, M. Hirano, H. Hosono,
%Light-induced conversion of an insulating refractory oxide 
%into a persistent conductor. 
{\it Nature} {\bf 2002}, {\it 419}, 462.
%-465 (2002).

\bibitem{may-science}
S. Matsuishi, Y. Toda, M. Miyakawa, K. Hayashi, T. Kamiya,
M. Hirano, T. Tanaka, H. Hosono, 
%High-density electron anions in a nanoporous single crystal: 
%[Ca$_{24}$Al$_{28}$O$_{64}$]$^{4+}$(4e$^{-}$).
{\it Science} {\bf 2003}, {\it 301}, 626.
% (2003)

\bibitem{PRLmay}
J.E. Medvedeva, A.J. Freeman, M.I. Bertoni, T.O. Mason, 
%Electronic structure and light-induced conductivity in a transparent refractory oxide.
{\it Phys. Rev. Lett.} {\bf 2004}, {\it 93}, 16408.
% (2004).

\bibitem{Minami}
T. Minami, 
%Transparent and conductive multicomponent oxide 
%films prepared by magnetron sputtering.
{\it J. Vac. Sci. Technol. A} {\bf 1999} {\it 17}, 1765.
%-1772 (1999).
%T. Minami, MRS Bull. {\bf 25}, 38 (2000).
%-44
%New n-type transparent conducting oxides.


\bibitem{FLAPW}
E. Wimmer, H. Krakauer, M. Weinert, A.J. Freeman, 
%Full-potential self-consistent linearized-augmented-plane-wave method 
%for calculating the electronic structure of molecules and surfaces -- O$_2$ molecule.
{\it Phys. Rev. B} {\bf 1981} {\it 24}, 864.
%-875 (1981).
\bibitem{FLAPW1}
M. Weinert, E. Wimmer, A.J. Freeman, 
%Total-energy all-electron density functional method 
%for bulk solids and surfaces.
{\it Phys. Rev. B} {\bf 1982}, {\it 26}, 4571.
%-4578 (1982).

\bibitem{str}
K. Kato, I. Kawada, N. Kimizuka, T. Katsura, 
%Die Kristallstructur von YbFe$_2$O$_4$.
{\it Z. Krist.} {\bf 1975}, {\it 141}, 314.
%-320 (1975).
\bibitem{str-all}
N. Kimizuka, T. Mohri, 
%Spinel, YbFe$_2$O$_4$, and Yb$_2$Fe$_3$O$_7$ types of structures for compounds in 
%the In$_2$O$_3$ and Sc$_2$O$_3$-A$_2$O$_3$-BO
%systems [A: Fe, Ga or Al; B: Mg, Mn, Fe, Ni, Cu or Zn] at temperatures over 1000 C.
{\it J. Solid State Chem.} {\bf 1985}, {\it 60}, 382.
%-384 (1985).

\bibitem{Li}
C. Li, Y. Bando, M. Nakamura, M. Kimizuka,
%A modulated structure of In$_2$O$_3$(ZnO)$_{(m)}$ as revealed by 
%high resolution electron microscopy.
{\it J. Electron Microsc.} {\bf 1997}, {\it 46}, 119.
%-127 (1997).

\bibitem{family}
N. Kimizuka, T. Mohri, 
%Structural classification of RAO$_3$(MO)$_n$ compounds 
%(R=Sc,In,Y or lanthanides; A=Fe(III),
%Ga,Cr, or Al; M=divalent cation; $n$=1-11).
{\it J. Solid State Chem.} {\bf 1989}, {\it 78}, 98.
%-107 (1989).

\end{thebibliography}
\end{document}